\documentclass{article}

\usepackage{graphicx}
\usepackage{times}
\usepackage{xcolor}
\usepackage{amsfonts}
\usepackage{esvect}
\usepackage{harpoon}
\usepackage{amsmath,accents}
\usepackage{braket}

\usepackage{cite}

\usepackage{subfigure,color}

\def\e{\begin{equation}}
\def\f{\end{equation}}
\def\_#1{{\bf #1}}
\def\=#1{\overline{\overline #1}}
\def\/#1{_{\rm #1}}
\def\.{\cdot}

\DeclareMathOperator\arctanh{arctanh}

\title{Quantum Theory of Wave Scattering from Electromagnetic Time Interfaces}

\newenvironment{sciabstract}{%
\begin{quote} \bf}
{\end{quote}}

\author
{M.~S.~Mirmoosa, T.~Set{\"a}l{\"a}, and A.~Norrman\\
\normalsize{Center for Photonics Sciences, University of Eastern Finland, P.O.~Box 111, FI-80101 Joensuu, Finland}
} 

\date{}

\begin{document} 

\baselineskip24pt

\maketitle 

%%%%%%%%%%%%%%%%%%%%%%%%%%%%%%%%%%%%%%%%%%%%%%%%%%%%%%%%%%%%%%%%%%%%%%%%%%%%%%% 

\begin{sciabstract}
{Modulating macroscopic parameters of materials in time offers innovative avenues for manipulating electromagnetic waves. Due to such enticing prospects, the general research subject of time-varying systems is expanding today in different branches of electromagnetism and optics. However, compared with the research efforts and progresses that have taken place in the realm of classical electrodynamics, the quantum aspects of this emerging subject have been less explored. Here, through the lens of quantum optics, we study the scattering of electromagnetic fields from an isotropic and nondispersive material with a suddenly changing refractive index, creating a time interface. We revisit the transformation of the bosonic mode operators and corresponding quantum states due to this interface, governed by the two-mode squeeze operator. Building on this foundation, more importantly, our analysis focuses on the photon statistics and quantum state engineering of the scattered light, elucidating various quantum optical phenomena and opportunities arising from the time interface. Notably, these include photon-pair production and destruction, photon bunching and antibunching, vacuum generation, quantum state discrimination, and quantum state freezing. To bridge theory and experiment, we propose a possible circuit quantum electrodynamics approach for validating our theoretical predictions. We hope that our work inspires experimental investigations and further quantum optical explorations of electromagnetic field interaction with photonic time crystals or with dispersive time-varying materials.}
\end{sciabstract} 

%%%%%%%%%%%%%%%%%%%%%%%%%%%%%%%%%%%%%%%%%%%%%%%%%%%%%%%%%%%%%%%%%%%%%%%%%%%%%%%

\maketitle

%%%%%%%%%%%%%%%%%%%%%%%%%%%%%%%%%%%%%%%%%%%%   Introduction   %%%%%%%%%%%%%%%%%%%%%%%%%%%%%%%%%%%%%%%%%%%%

\section{Introduction} 

Investigation of field interaction with materials whose effective parameters vary in time is currently at the core of research in the electrodynamics community~\cite{galiffi2022photonics, Ptitcyn2023Tutorial}. This is because such a temporal variation renders an additional degree of freedom for controlling electromagnetic waves in a desired way~\cite{engheta2023four,Engheta20NPH}. One of the basic scenarios within this area is to consider a temporal discontinuity according to which the effective parameters suddenly change from one value to another, although they remain uniform in space~\cite{morgenthaler1958velocity}. In analogy to a spatial discontinuity, this results in the emergence of reflected waves, and, in contrast, it gives rise to the frequency translation phenomenon and the breaking of the conservation of power~\cite{morgenthaler1958velocity,mendoncca2002time,Agrawal2014RTC}. Besides such important occurrences which were studied theoretically~\cite{morgenthaler1958velocity,Wilks1988TH,mendoncca2002time,Agrawal2014RTC} as well as experimentally~\cite{Yugami2002EXPER,nishida2012EXP,water,Alu_TL}, other captivating effects have been reported. Indeed, by analyzing temporal discontinuities in isotropic, anisotropic, and bianisotropic materials, various possibilities, applications, and effects have been uncovered including the realization of dispersion bands separated by wavevector gaps~\cite{Segev8PTC}, anti-reflection temporal coatings~\cite{ramaccia2020light,pacheco2020anti}, inverse prism~\cite{akbarzadeh2018inverse}, temporal aiming~\cite{pacheco2020aiming}, polarization conversion~\cite{xu2021complete}, polarization splitting~\cite{mostafa2023spin}, polarization-dependent analog computing~\cite{RizzaCastaldiGaldi23}, polarization rotation and direction-dependent wave manipulation~\cite{mirmoosa2023timeIIBM}, wave freezing and melting~\cite{wang2023controlling}, transformation of surface waves into free-space radiation~\cite{Grap19SPP,wang2023controlling}, and so forth. 

Beyond classical electrodynamics, the subject of temporal discontinuities in materials has engrossed attention also in quantum optics~\cite{mendoncca2000quantum, SecondRVAgree1, SecondRVAgree2, vazquez2022shaping,liberal2023quantum}. However, until today, the research efforts in this domain have been rather limited as compared with those in classical electrodynamics. Thus, exploring and understanding the basic quantum optical principles of temporal interfaces is important not only for advancing the research area on time-varying materials but also for the development of quantum photonics. In this paper, we present a detailed investigation of quantum light scattering at an electromagnetic time interface between two isotropic and nondispersive media, with considerations including photon statistics and quantum state engineering. In particular, by considering the case in which a forward and a backward propagating mode exist before the temporal discontinuity, we show that the time interface leads to a unitary evolution of the bosonic mode operators and the corresponding quantum states in terms of the two-mode squeeze operator (as indicated also by Ref.~\cite{mendoncca2000quantum}). Then, we examine in detail several quantum optical features related to the output state when the initial forward propagating mode is in a number state and the backward propagating mode is in the vacuum state. More precisely, we elucidate the photon probability distribution and demonstrate that photon-pair generation is a fundamental inherent feature of the temporal interface. We analyze various conditions for such photon-pair generation, including the one under which the maximum probability for generating a one-photon pair out of initial vacuum occurs. In addition, importantly, we study the photon number fluctuations as well as the degree of second-order coherence of the individual output modes. It is shown that the time interface creates noise such that the backward propagating mode after the temporal discontinuity follows super-Poissonian photon statistics and corresponds to bunched light regardless of the refractive indices of the two media. On the other hand, the photon statistics of the forward propagating mode can be tuned by the refractive-index ratio between sub-Poissonian (antibunched) and super-Poissonian (bunched). We revisit the salient features above about photon statistics when the initial state is a coherent state instead of a number state. In this case, both of the output modes are verified to display super-Poissonian (bunched) light character.  

Furthermore, in this paper, we demonstrate also how challenging our preliminary assumptions can unveil interesting phenomena. First, by introducing photons in the initial backward propagating mode as well, we reveal that photon-pair destruction, alongside photon-pair production, is also a central property of the temporal interface. We illustrate how this phenomenon enables the simultaneous generation of vacuum states for both of the output modes and facilitates quantum state discrimination. Second, by extending our analysis to materials with magnetic responses, we uncover the possibility of preserving the initial quantum state. Lastly, we suggest an experimental approach utilizing superconducting transmission lines operating at extremely low temperatures to empirically validate the theoretical findings presented in this work.

The paper is organized as follows: In Section~\ref{sec:classical}, we briefly describe the concept of a temporal discontinuity based on classical electromagnetics, which facilitates the subsequent quantum description. In Section~\ref{sec:QuantPic}, which is the primary contribution of this paper, we study the case from the quantum optics perspective as explained above. Finally, in Section.~\ref{sec:conclusion}, we summarize the main conclusions of the paper and discuss future outlooks and prospects of this research direction.   

It is worth mentioning that recently, in parallel with our current work, several papers have appeared which belong to the broader area of quantum field interaction with time-varying materials. These include the quantum treatment of modulations moving at a uniform speed~\cite{RVNotAgree1, RVNotAgree2}, temporal modulations with an arbitrary profile~\cite{RVNotAgree3}, and periodic modulations~\cite{RVNotAgree4}.

%%%%%%%%%%%%%%%%%%%%%%%%%%%%%%%%%%%%%%%%%%%   Classical Picture
%%%%%%%%%%%%%%%%%%%%%%%%%%%%%%%%%%%%%%%%%%%  

\section{Classical Fields}
\label{sec:classical}

\begin{figure}[!t]
\centerline{\includegraphics[width=1\columnwidth]{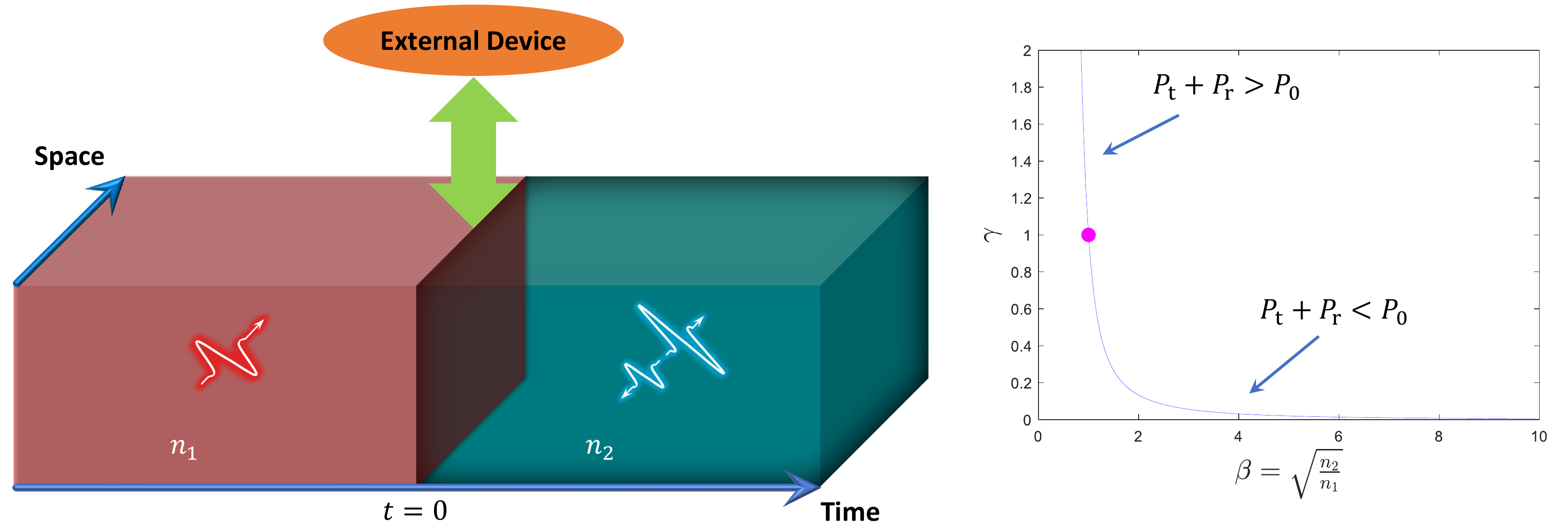}}
\caption{Time interface in a dielectric medium. Here, $n_1$ and $n_2$ denote the refractive index before and after the temporal jump ($t=0$), respectively. There is an external device which interacts with the system in order to change the refractive index at $t=0$. As a consequence, the initial wave ($t<0$, red signal) splits into reflected and transmitted waves with angular frequency translation ($t>0$, blue signals). Additionally, the ratio between the scattered and incident averaged powers $\gamma=(P_{\rm{t}}+P_{\rm{r}})/P_0$ as a function of the refractive-index ratio $\beta=\sqrt{n_2/n_1}$ is also shown, illustrating the breaking of the conservation of power.}
\label{fig:timeORG}
\end{figure}

Consider an abrupt (step-like) temporal change of the refractive index of a homogeneous nondispersive linear dielectric medium. As shown in Fig.~\ref{fig:timeORG}, the refractive index before and after the temporal jump at $t=0$ is denoted by $n_1$ and $n_2$, respectively. This kind of change called a temporal discontinuity, temporal boundary, or time interface is in contrast to a spatial boundary where the refractive index is discontinuous in space but uniform in time. As an example, we consider an initial plane wave which is propagating in the $z$-direction and is polarized along the $x$-axis. Hence, regarding $t<0$, the real electric field is written as $\_E(z,t<0)=\mathrm{Re}(E_0e^{ikz}e^{-i\omega_1t})\_a_x$, where $E_0$ is the complex field amplitude, $\omega_1$ denotes the angular frequency, $k$ is the corresponding wavenumber, and $\_a_x$ represents the unit vector along the positive $x$-axis in the Cartesian coordinate system. Due to the presence of the time interface, the electric field for $t>0$ becomes a superposition of two plane waves with different amplitudes ($E_{\rm{r}}$ and $E_{\rm{t}}$) and angular frequencies ($\omega_r$ and $\omega_t$):
\begin{equation}
\_E(z,t>0)=\mathrm{Re}\big(E_{\rm{r}}e^{ik z}e^{-i\omega_rt}+E_{\rm{t}}e^{ik z}e^{-i\omega_tt}\big)\_a_x.
\label{eq:EQEFTCP}
\end{equation}
Since the medium does not vary in space, the wavenumber of the two plane waves in Eq.~(\ref{eq:EQEFTCP}) equals to that of the initial wave due to the conservation of linear momentum~\cite{galiffi2022photonics}. This property leads to the dispersion relation 
\begin{equation}
k^2={\omega_1^2\over c^2}n_1^2={\omega_r^2\over c^2}n_2^2={\omega_t^2\over c^2}n_2^2, 
\end{equation}
with $c$ being the speed of light in a vacuum, from which we conclude that 
\begin{equation}
\omega_t={n_1\over n_2}\omega_1,\quad\omega_t=-\omega_r.
\label{eq:eq:AFCCP}
\end{equation} 
Note that the second relation in Eq.~(\ref{eq:eq:AFCCP}) implies that $\omega_t$ and $\omega_r$ cannot be equal. In particular, because $\omega_r<0$, this angular frequency refers to a plane wave whose phase velocity is in the opposite direction compared to the one with $\omega_t>0$. Thus, these two plane waves correspond to the backward and forward waves. From a mathematical point of view, a plane wave with negative angular frequency and positive wavenumber is transferring energy in the {\it{positive-index}} medium quite similarly to a plane wave with positive angular frequency and negative wavenumber. This interesting characteristic enforces the postulation that the generated backward wave is in fact a reflected wave in space. 

%%%%%%%%%%%%%%%%%%%% Figure %%%%%%%%%%%

%%%%%%%%%%%%%%%%%%%%%%%%%%%%%%%%%%%%

Having forward (transmitted) and backward (reflected) waves, we can define transmission and reflection coefficients. For deriving them, similarly to what we do concerning spatial boundaries, we need boundary conditions. Since in Maxwell's equations there are time derivatives of the electric and magnetic flux densities ($\_D$ and $\_B$), these two vectors should be continuous at the vicinity of $t=0$ when the refractive index abruptly increases or decreases (see, e.g., Ref.~\cite[Sec.~2]{galiffi2022photonics}). Hence, on approaching the temporal boundary from the negative $(t=0^-)$ and positive $(t=0^+)$ time directions, the following important conditions must hold:
\begin{equation}
\begin{split}
&\_D(z,t=0^-)=\_D(z,t=0^+),\cr
&\_B(z,t=0^-)=\_B(z,t=0^+).
\end{split}
\label{eq:EQBCCPDB}
\end{equation}
It is evident that if the medium does not have a magnetic response, the lower row in the above equation means the continuity of the magnetic field ($\_H$). In the case of locality in time, the electric flux density is connected to the electric field as $\_D(z,t)=\epsilon\_E(z,t)$, in which $\epsilon/\epsilon_0=n^2$ is the relative permittivity ($\epsilon_0$ denotes the vacuum permittivity). Based on these assumptions, we eventually deduce from Eqs.~\eqref{eq:EQEFTCP} and \eqref{eq:EQBCCPDB} as well as Maxwell's equations that the reflection ($R$) and transmission ($T$) coefficients depend on the refractive indices according to 
\begin{equation}
R={E_{\rm{r}}\over E_0}={n_1\over2n_2}\Big({n_1\over n_2}-1\Big),\quad 
T={E_{\rm{t}}\over E_0}={n_1\over2n_2}\Big({n_1\over n_2}+1\Big).
\label{eq:eq:RTTR}
\end{equation} 
Equation~\eqref{eq:eq:RTTR} indicates correctly that if $n_1=n_2$ the reflection becomes zero, and we have full transmission. 

Finally, it is worth mentioning about the conservation of power. In the case of a spatial boundary, the ratio between the summation of reflected ($P_{\rm{r}}$) and transmitted ($P_{\rm{t}}$) averaged power per unit area (or intensity) and the incident ($P_0$) averaged power per unit area is unity. Indeed, $\gamma=(P_{\rm{t}}+P_{\rm{r}})/P_0=R_{\rm{s}}^2+(n_2/n_1)T_{\rm{s}}^2=1$, in which the reflection and transmission coefficients are given by $R_{\rm{s}}=(n_1-n_2)/(n_1+n_2)$ and $T_{\rm{s}}=2n_1/(n_1+n_2)$, respectively~\cite{cheng1989field} (the subscript ``s" refers to ``spatial"). However, for the time interface, we instead find that $\gamma=(P_{\rm{t}}+P_{\rm{r}})/P_0=(n_2/n_1)(R^2+T^2)$, which according to Eq.~\eqref{eq:eq:RTTR} equals to 
\begin{equation}
\gamma={n_1\over2n_2}\Bigg[\bigg({n_1\over n_2}\bigg)^2+1\Bigg].
\label{eq:eq:RCPPTPR}
\end{equation} 
Thus, since Eq.~(\ref{eq:eq:RCPPTPR}) never goes to unity except when $n_1=n_2$, the conservation of power does not hold for the time interface. This is because there is a pumping (or modulating) system which needs to do work in order to change the refractive index in time. Due to this work, the summation of the reflected and transmitted averaged power densities cannot be the same as the averaged power density associated with the incident wave before $t=0$. Notice that the ratio $\gamma$ can be smaller or larger than unity depending on the refractive-index ratio, as shown by Eq.~\eqref{eq:eq:RCPPTPR}.              

In summary, from the classical point of view, we succinctly explained the most salient features of a temporal discontinuity: angular frequency conversion [Eq.~\eqref{eq:eq:AFCCP}], creation of reflected waves [Eq.~\eqref{eq:eq:RTTR}], and breaking the conservation of power [Eq.~\eqref{eq:eq:RCPPTPR}]. Remembering these three properties which have been also illustrated in Fig.~\ref{fig:timeORG}, in the following, we move to examine the quantum aspects of the fields.

%%%%%%%%%%%%%%%%%%%%%%%%%%%%%%%%%%%%%%%%%%%%%%%%%%%%%%%%%  Section 3  %%%%%%%%%%%%%%%%%%%%%%%%%%%%%%%%%%%%%%%%%%%%%%%%%%%%%%%%%

\section{Quantum Fields} 
\label{sec:QuantPic} 
This is the leading part of our study, focusing on the scattering of electromagnetic fields from the temporal interface within a quantum optics framework. As a first step, we derive the unitary operator which relates the mode operators and the quantum states before and after the time interface. Based on this central derivation, we continue by investigating photon-pair generation via the probability distribution of the output state as well as the photon statistics of the output modes when the initial state is in a number state. Also, we revisit the analysis of photon statistics when the initial state is in a coherent state. Subsequently, we elucidate several important phenomena--photon-pair destruction, vacuum generation, quantum state discrimination, and quantum state freezing--all of which are achieved by relaxing certain initial assumptions. At the end of the section, we suggest an approach that enables us to verify the theoretical predictions introduced in this work.

%Thus, as the first step, we deduce the output quantum states for the input Fock state that is necessarily sub-Poissonian and anti-bunched with zero variance for the number operator. Notice that at the end of the current section, we revisit our inspections for the case of a coherent state as the input one, and, therefore, we see the differences in output results corresponding to these two input states. 

%%%%%%%%%%%%%%%%%%%%%%%%%%%%%%%%%%%%%%%%%%%%%%

\subsection{Transformation of mode operators}

We start by writing the electric field operator. Unlike in the previous classical treatment, we now extend our analysis to the case where two modes (forward and backward) exist simultaneously before ($t<0$) and after ($t>0$) the temporal jump at $t=0$. The propagation directions of these forward and backward modes are characterized by the arbitrary wavevectors $\_k$ and $-\_k$, respectively. In addition, we consider an arbitrary polarization direction, specified by the unit vector $\_e$, which is orthogonal to the propagation axis. Consequently, the electric field operator $\hat{\_E}(\_r,t)$ at position $\_r$ is expressed as~\cite{loudon2000quantum} 
\begin{equation}
\begin{split}
&\hat{\_E}(\_r,t<0)=i\Big(\frac{\hbar\omega_{1}}{2\epsilon_1 V}\Big)^{1/2}\sum_{\_k^\prime}\big[\hat{a}_{\_k^\prime}e^{i(\_k^\prime\cdot\_r-\omega_{1}t)}-\hat{a}_{\_k^\prime}^\dagger e^{-i(\_k^\prime\cdot\_r-\omega_{1}t)}\big]\_e,\cr
&\hat{\_E}(\_r,t>0)=i\Big(\frac{\hbar\omega_{2}}{2\epsilon_2 V}\Big)^{1/2}\sum_{\_k^\prime}\big[\hat{b}_{\_k^\prime}e^{i(\_k^\prime\cdot\_r-\omega_{2}t)}-\hat{b}_{\_k^\prime}^\dagger e^{-i(\_k^\prime\cdot\_r-\omega_{2}t)}\big]\_e.
\end{split}
\label{EQ:EQBAVFOEVBAV}
\end{equation} 
Here, $\hbar$ is the reduced Planck constant, $\omega_1$ and $\epsilon_1$ ($\omega_2$ and $\epsilon_2$) are the angular frequency and permittivity for $t<0$ ($t>0$), $V$ is the quantization volume, and $\_k^\prime\in\{\_k,-\_k\}$. Moreover, and in particular, $(\hat{a},\,\hat{a}^\dagger)$ and $(\hat{b},\,\hat{b}^\dagger)$ denote the bosonic annihilation and creation operators before and after $t=0$, respectively. These operators are connected to each other via the electromagnetic boundary conditions which state that the flux density operators must be continuous at $t=0$. In other words, quite similarly to the classical picture [Eq.~(\ref{eq:EQBCCPDB})], we have
\begin{equation}
\begin{split}
&\hat{\_D}(\_r,t=0^-)=\hat{\_D}(\_r,t=0^+),\cr
&\hat{\_B}(\_r,t=0^-)=\hat{\_B}(\_r,t=0^+).
\label{eq:QBCS}
\end{split}
\end{equation}
The electric flux density is readily calculated because $\hat{\_D}=\epsilon\hat{\_E}$ (assuming temporal locality~\cite{mirmoosa2022dipole} as before), whereas the magnetic flux density is obtained from Faraday's law $\nabla\times\hat{\_E}=-{\partial\hat{\_B}/\partial t}$. Eventually, by imposing the boundary conditions in Eq.~\eqref{eq:QBCS} and doing some algebraic manipulations, we conclude that
\begin{equation}
\begin{pmatrix}
\hat{b}_{\_k} \\
\hat{b}_{-\_k}^\dagger
\end{pmatrix}=
\begin{pmatrix}
\tau & \Gamma \\
\Gamma & \tau
\end{pmatrix}
\!
\begin{pmatrix}
\hat{a}_{\_k}\\
\hat{a}_{-\_k}^\dagger
\end{pmatrix},
\label{eq:bamodes1}
\end{equation}
where we introduced the two quantities
\begin{equation}
\tau=\frac{1}{2}\Big(\sqrt{\frac{n_2}{n_1}}+\sqrt{\frac{n_1}{n_2}}\Big), \quad
\Gamma=\frac{1}{2}\Big(\sqrt{\frac{n_2}{n_1}}-\sqrt{\frac{n_1}{n_2}}\Big).
\label{eq:taugamma}
\end{equation}
Equations~(\ref{eq:bamodes1}) and (\ref{eq:taugamma}) form a central result of our work. They show that the temporal discontinuity is a transformation by which the annihilation operator $\hat{b}_{\_k}$ (creation operator $\hat{b}^\dagger_{-\_k}$) of the output forward (backward) mode becomes a mixture of the annihilation (creation) and creation (annihilation) operators of the input forward (backward) and backward (forward) modes, respectively.

To obtain more insight into these relations, let us have a closer look at Eq.~(\ref{eq:taugamma}). First, observing that $\tau+\Gamma=\sqrt{n_2/n_1}$ and $\tau-\Gamma=\sqrt{n_1/n_2}$, we conclude that $\tau+\Gamma$ is the inverse of $\tau-\Gamma$. Second, and more importantly,    
\begin{equation}
\tau^2-\Gamma^2=1,
\label{eq:TauGammaR}
\end{equation} 
which allows us to parametrize $\tau$ and $\Gamma$ via hyperbolic functions, i.e., $\tau=\cosh r$ and $\Gamma=\sinh r$. Owing to such a representation, we rewrite Eq.~(\ref{eq:bamodes1}) as 
\begin{equation}
\begin{pmatrix}
\hat{b}_{\_k} \\
\hat{b}_{-\_k}^\dagger
\end{pmatrix}=
\begin{pmatrix}
\cosh r & \sinh r\\
\sinh r & \cosh r
\end{pmatrix}
\!
\begin{pmatrix}
\hat{a}_{\_k}\\
\hat{a}_{-\_k}^\dagger
\end{pmatrix}.
\label{eq:bamodes2}
\end{equation} 
Very interestingly, from this form, we discover that the time interface acts as a particular Bogoliubov transformation between the mode operators, as in the context of optical parametric amplification and the Unruh effect~\cite{leonhardt2010essential}. Furthermore, it is clear that the transformation matrix in Eq.~\eqref{eq:bamodes2} is real and symmetric, having a determinant equal to one. However, we see that the matrix is not orthogonal, and hence not unitary, which is a reflection of energy nonconservation owing to the external work required to change the refractive index.

%%%%%%%%%%%%%%%%%%%%%%%%%%%%%%%%%%%%%%%%%%%%%%%%%%%%%%%%%%%%%%

The Bogoliubov transformation in Eq.~(\ref{eq:bamodes2}) advocates that the unitary evolution of the modes is governed by the two-mode squeeze operator~\cite{agarwal2012quantum,gerry2023introductory} 
\begin{equation}
\hat{S}(r)=e^{r(\hat{a}_\_k\hat{a}_{-\_k}-\hat{a}^\dagger_\_k\hat{a}^\dagger_{-\_k})}.
\label{eq:Soperator}
\end{equation}
Indeed, by employing the Baker-Campbell lemma~\cite{sakurai2020modern}, commutators of the $\hat{a}$ operators, and Taylor series of hyperbolic functions, we verify from Eqs.~(\ref{eq:bamodes2}) and (\ref{eq:Soperator}) that
\begin{equation}
\hat{S}(r)
\begin{pmatrix}
\hat{a}_{\_k} \\
\hat{a}_{-\_k}^\dagger
\end{pmatrix}
\hat{S}^\dagger(r)=
\begin{pmatrix}
\hat{b}_{\_k} \\
\hat{b}_{-\_k}^\dagger
\end{pmatrix}.
\label{EQUT:EM12NOVUTC}
\end{equation}
Furthermore, from Eqs.~(\ref{eq:bamodes1}), (\ref{eq:taugamma}), and (\ref{eq:bamodes2}), we find that the squeezing parameter $r$ is purely real and determined by the refractive indices according to
\begin{equation}
r=\arctanh{\Big(\frac{\Gamma}{\tau}\Big)}=\arctanh{\Big(\frac{n_2-n_1}{n_2+n_1}\Big)}.
\label{EQEQEQRRR}
\end{equation}
In other words, $\hat{b}$ is the unitary transformation of $\hat{a}$ and, accordingly, $\hat{S}(r)$ corresponds to the unitary evolution operator $\hat{U}(t)=\exp(-{i\over\hbar}\hat{H}t)$, where $\hat{H}=i\hbar g(\hat{a}_\_k\hat{a}_{-\_k}-\hat{a}^\dagger_\_k\hat{a}^\dagger_{-\_k})$ is the Hamiltonian and $gt=r$.

%%%%%%%%%%%%%%%%%%%%%%%%%%%%%%%%%%%%%%%%%%%%%%%%%%%%%%%%%%%%%%

\subsection{Output quantum states}

Having derived the mode transformation matrix and the associated unitary evolution operator for the time interface, we now turn to investigate how the initial quantum state and the underlying photon properties are altered due to the temporal jump. To this end, we examine in detail the scenario where the initial state $(t<0)$ is a Fock state $\ket{n,0}$. This means that the forward incident mode contains $n$ photons and the backward incident mode is in the vacuum state. The output state $\ket{\psi}$ $(t>0)$ is found by applying the two-mode squeeze operator $\hat{S}(r)$ in Eq.~(\ref{eq:Soperator}) to the initial state, viz., $\ket{\psi}=\hat{S}(r)\ket{n,0}$. For this purpose, because the $\hat{S}(r)$ operator is in the form of an exponential, it is convenient to employ the disentanglement theorem~\cite{wodkiewicz1985coherent} and rewrite it as     
\begin{equation}
\hat{S}(r)=e^{\tanh(r)\hat{K}_+}e^{\ln[1-\tanh^2(r)]\hat{K}_0}e^{-\tanh(r)\hat{K}_-}, 
\label{eq:TERMs}
\end{equation}
where $\hat{K}_-=\hat{a}_\_k\hat{a}_{-\_k}$, $\hat{K}_+=\hat{a}^\dagger_\_k\hat{a}^\dagger_{-\_k}$, and $\hat{K}_0=(1/2)(\hat{a}^\dagger_\_k\hat{a}_\_k+\hat{a}^\dagger_{-\_k}\hat{a}_{-\_k}+1)$. These Hermitian operators obey the commutation relations $[\hat{K}_0,\,\hat{K}_\pm]=\pm\hat{K}_\pm$ and $[\hat{K}_-,\,\hat{K}_+]=2\hat{K}_0$, so they define an SU(1,1) algebra~\cite{gilles1992non}. We stress that the decomposition in Eq.~(\ref{eq:TERMs}) is a general property of the two-mode squeeze operator, i.e., it is independent of our chosen initial state $\ket{n,0}$. 

The expression for $\hat{S}(r)$ in Eq.~\eqref{eq:TERMs} involves three terms. The last term preserves the initial state, because it contains only annihilation operators and the backward incident mode is in the vacuum state. Regarding the middle term, we obtain that 
\begin{equation}
e^{\ln[1-\tanh^2(r)]\hat{K}_0}\ket{n,0}=[1-\tanh^2(r)]^{n+1\over2}\ket{n,0}.
\end{equation} 
In other words, $[1-\tanh^2(r)]^{n+1\over2}$ is the eigenvalue of the operator in the middle with the eigenstate $\ket{n,0}$. In consequence, we observe that the last and middle terms do not change the state and $\ket{n,0}$ is conserved. However, the situation is drastically different for the first term because it includes the creation operators. Indeed, if we apply the standard operator identities $(\hat{a}^\dagger)^m\ket{0}=\sqrt{m!}\ket{m}$ and $\hat{a}^\dagger\ket{m}=\sqrt{m+1}\ket{m+1}$~\cite{GriffithsIQM}, we infer the output state to be~\cite{gilles1992non}
\begin{equation}
\ket{\psi}=\big[1-\tanh^2(r)\big]^{1+n\over2}\sum_{m=0}^\infty\sqrt{{(n+m)!\over n!m!}}\tanh^m(r)\ket{n+m,m}.
\label{eq:eq:OSAJ1}
\end{equation}
This expression explicitly demonstrates that the first term in Eq.~\eqref{eq:TERMs} transforms the initial state into an entangled superposition state of higher photon numbers specified by $m$. In terms of the material parameters $\Gamma=\sinh r$ and $\tau=\cosh r$, we find that Eq.~(\ref{eq:eq:OSAJ1}) takes the form
\begin{equation}
\ket{\psi}=\Big(\frac{1}{\tau}\Big)^{1+n}\sum_{m=0}^\infty\sqrt{{(n+m)!\over n!m!}}\Big(\frac{\Gamma}{\tau}\Big)^m\ket{n+m,m}.
\label{eq:eq:OSAJ2}
\end{equation}
Thus, the time interface gives rise to photon-pair generation, and the initial state $\ket{n,0}$ is only a part of the distribution of the output state $\ket{\psi}$ that contributes with the probability of $[1-\tanh^2(r)]^{1+n}=(1/\tau^2)^{1+n}$. Photon-pair generation, characterized by the integer $m$ in Eqs.~\eqref{eq:eq:OSAJ1} and \eqref{eq:eq:OSAJ2}, serves as a quantum optical analogue to the time reflection phenomenon observed in classical wave dynamics. At a more fundamental level, this process is in fact rooted in the principle of {\it wavenumber conservation} (or {\it linear momentum conservation}), which governs the creation of a negative frequency component, as detailed in Section~\ref{sec:classical} [see, Eq.~\eqref{eq:eq:AFCCP}]. Remember that in the classical picture, the concept of negative frequency corresponds to time reflection.

When not only the backward incident mode but also the forward incident mode is in the vacuum state ($n=0$), Eq.~(\ref{eq:eq:OSAJ1}) shows that the output state becomes the two-mode squeezed vacuum~\cite{agarwal2012quantum,gerry2023introductory}:
\begin{equation}
\ket{\psi}=\frac{1}{\cosh r}\sum_{m=0}^\infty\tanh^m(r)\ket{m,m}=\frac{1}{\tau}\sum_{m=0}^\infty\Big(\frac{\Gamma}{\tau}\Big)^m\ket{m,m}.
\label{eq:vacuum}
\end{equation}
In this case merely the paired states $\ket{m,m}$ appear in the superposition, which makes Eq.~(\ref{eq:vacuum}) the strongest entangled state of two light modes with given total energy~\cite{barnett1989entropy,barnett1991information}. After the interface, our system consists of two distinguishable but correlated subsystems: The output forward and backward propagating modes that are distinguished by opposite propagation directions. Due to the superposition in Eq.~\eqref{eq:vacuum}, we do not precisely know the number of photons in each mode. However, upon measuring the photon count in one mode, we instantaneously ascertain the number of photons in the other, highlighting the quantum entanglement between these two modes.

%In fact, by comparing this matrix with the one written for the Bogoliubov transformation, we conclude that the temporal discontinuity is associated with a bosonic two-mode Bogoliubov transformation.

%%%%%%%%%%%%%%%%%%%%%%%%%%%%%%%%%%%%%%%%%%%%%%%%%%%%%%%%%%%%%%%%%%%%%%%%%%%%%%

\subsection{Probability distribution and photon-pair generation}

\begin{figure}[!t]
\centerline{\includegraphics[width=0.5\columnwidth]{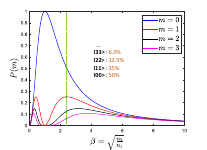}}
\caption{The probability distribution $P(m)$ as a function of the refractive-index ratio $\beta=\sqrt{n_2/n_1}$ for different values of excited photons $m$ when the initial state is $\ket{0,0}$. The small inset in the center of the figure indicates probabilities associated with $\vert m\,m\rangle$ ($m=0,1,2,3$) under the condition that the possibility of one photon-pair generation is maximum. The vertical green-dotted line and the solid green circles on it specify the locations that correspond to such probabilities shown in the aforementioned inset.}
\label{fig:POO00}
\end{figure}

According to Eqs.~\eqref{eq:eq:OSAJ1} and \eqref{eq:eq:OSAJ2}, the probability-distribution function of the output state is
\begin{equation}
\begin{split}
P(n,m)&=\big[1-\tanh^2(r)\big]^{1+n}\tanh^{2m}(r){(n+m)!\over n!m!} \cr
&=\Big(\frac{1}{\tau^2}\Big)^{n+1}\Big(\frac{\Gamma}{\tau}\Big)^{2m}\frac{(n+m)!}{(n!m!)}.
\label{eq:probabilityP}
\end{split}
\end{equation}
From the theory of series, we know that $1/(1-x)^{n+1}=\sum_mx^m(n+m)!/(n!m!)$ for $\vert x\vert<1$. Replacing $x$ by $(\Gamma/\tau)^2$ we confirm that $\sum_m{P(n,m)}=1$, corroborating that $P(n,m)$ is a proper probability-distribution function. Let us inspect the scenario in which the initial state is a vacuum state $(n=0)$ and, as a result, the output state is in the two-mode squeezed vacuum as given by Eq.~\eqref{eq:vacuum}. Accordingly, the expression in Eqs.~\eqref{eq:probabilityP} is reduced to 
\begin{equation}
P(m)={\sinh^{2m}(r)\over\cosh^{2m+2}(r)}={\Gamma^{2m}\over\tau^{2m+2}}=4\beta^2{(\beta^2-1)^{2m}\over(\beta^2+1)^{2m+2}},
\label{eq:ISVSGS}
\end{equation}
where $\beta=\sqrt{n_2/n_1}$. We observe from Eq.~\eqref{eq:ISVSGS} that the probability of remaining in the vacuum state $m=0$ after the temporal jump is $P(0)=4\beta^2/(\beta^2+1)^2$. In particular, when $\beta\rightarrow0$ (which corresponds to changing to a zero-index material or low-index material compared to the initial medium) or when $\beta$ is very large (jumping to a high-index material) such possibility is negligible. To create a one-photon pair ($m=1$), the corresponding probability equals to $P(1)=4\beta^2(\beta^2-1)^2/(\beta^2+1)^4$. This probability is smaller than for the vacuum state ($m=0$), but it can still be reasonably large for specific values of $\beta$. The maximum of $P(1)$ is found where the derivative with respect to $\beta$ vanishes, i.e., $\beta=\sqrt{3\pm2\sqrt{2}}$ or $n_2=(3\pm2\sqrt{2})n_1$. In this scenario, the probability of generating a one-photon pair is maximally 25$\%$ for both positive and negative signs. On substituting these particular values of $\beta$ into the expression for $P(0)$ we find that there is a 50{\%} chance to remain in the vacuum state (regardless of those signs), which is only 2 times larger than for $P(1)$. These results are illustrated in Fig.~\ref{fig:POO00} in which the probability function $P(m)$ described by Eq.~\eqref{eq:ISVSGS} is plotted as a function of $\beta$ for different values of $m$. The figure shows that for $\beta=1$ the probability $P(0)=1$, and it is zero for $m\neq0$. Figure~\ref{fig:POO00} also shows that when the value of $\beta$ is large enough, the probability distribution becomes more uniform regarding different excited states. Indeed, if $\beta\gg1$, Eq.~\eqref{eq:ISVSGS} reduces to $P(m)\approx4/\beta^2$, confirming that the probability function becomes independent of $m$.

\begin{figure*}[!t]
\centerline{\includegraphics[width=0.9\columnwidth]{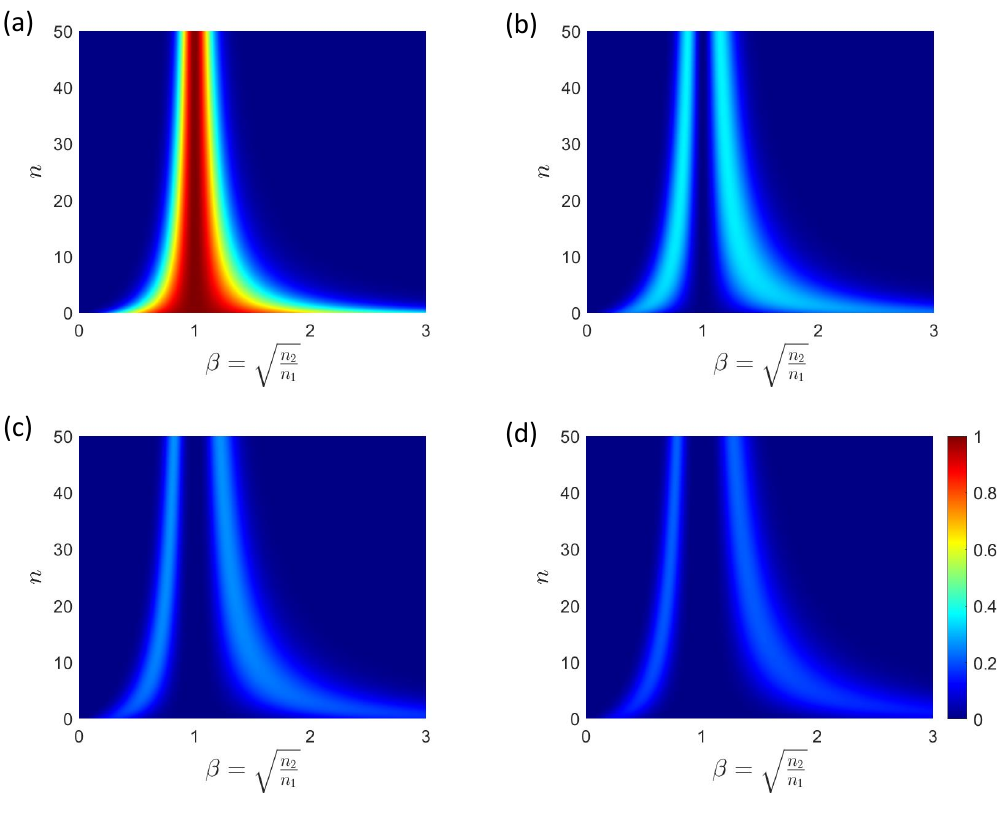}}
\caption{The probability distribution $P(n,m)$ with respect to the initial photon number $n$ and refractive-index ratio $\beta=\sqrt{n_2/n_1}$ for selected values of excited photons $m$. (a)~$m=0$, (b)~$m=1$, (c)~$m=2$, and (d)~$m=3$.}
\label{fig:PDFM}
\end{figure*}

The probability distribution with respect to $n$ and $\beta$ for several values of $m$ is presented in Fig.~\ref{fig:PDFM}. We see from Fig.~\ref{fig:PDFM}(a) that the probability of no photon creation ($m=0$) is high when the value of $\beta$ is close to unity. However, when $\beta$ deviates from unity, the probability $P(n,0)$ decreases. At the same time, the probabilities $P(n,1)$, $P(n,2)$, and $P(n,3)$ to excite one-, two-, and three-photon pairs increase and even surpass $P(n,0)$ for certain values of $n$, as indicated by Figs.~\ref{fig:PDFM}(b)--\ref{fig:PDFM}(d). Regarding the generation of a one-photon pair, Fig.~\ref{fig:PDFM}(b) illustrates that there is the possibility of roughly 35{\%} to achieve $m=1$. The exact maximum for a given $n$ is obtained analytically by considering the derivative with respect to $\beta$ of       
\begin{equation}
P(n,1)=\Big(\frac{\beta^2-1}{\beta^2+1}\Big)^2(n+1)\bigg[1-\Big(\frac{\beta^2-1}{\beta^2+1}\Big)^2\bigg]^{n+1}.
\end{equation}
Accordingly, we deduce that $[(\beta^2-1)/(\beta^2+1)]^2=1/(n+2)$ which results in $P_{\rm{max}}(n,1)=[(n+1)/(n+2)]^{n+2}$. The lower limit of $25{\%}$ is associated with $n=0$, as we discussed below Eq.~\eqref{eq:ISVSGS}, and the upper limit of $36{\%}$ corresponds to high values of $n$ [recall that $(1-x)^{1/x}=1/e\approx0.36$ when $x$ approaches zero].

%%%%%%%%%%%%%%%%%%%%%%%%%%%%%%%%%%%%%%%%%%%%%%%%%%%%%%%%%%%%%%%%%%%%%%

\subsection{Photon number fluctuations} 

Next, we assess the photon statistics of the forward and backward output modes by studying the expectation values and variances of their corresponding number operators $\hat{n}_{b_{\_k}}=\hat{b}_{\_k}^\dagger\hat{b}_{\_k}$ and $\hat{n}_{b_{-\_k}}=\hat{b}_{-\_k}^\dagger\hat{b}_{-\_k}$. According to Eq.~(\ref{eq:bamodes1}),
\begin{equation}
\begin{split}
\hat{n}_{b_{\_k}}&=\tau^2\hat{a}_{\_k}^\dagger\hat{a}_{\_k}+\Gamma^2\hat{a}_{-\_k}\hat{a}_{-\_k}^\dagger+\Gamma\tau\big(\hat{a}_{\_k}^\dagger\hat{a}_{-\_k}^\dagger+\hat{a}_{-\_k}\hat{a}_{\_k}\big),\cr 
\hat{n}_{b_{-\_k}}&=\Gamma^2\hat{a}_{\_k}\hat{a}_{\_k}^\dagger+\tau^2\hat{a}_{-\_k}^\dagger\hat{a}_{-\_k}+\Gamma\tau\big(\hat{a}_{\_k}\hat{a}_{-\_k}+\hat{a}_{-\_k}^\dagger\hat{a}_{\_k}^\dagger\big).
\end{split}
\label{eq:NOPBTD}
\end{equation}
Because the initial state before the temporal jump is taken to be $\ket{n,0}$, Eq.~\eqref{eq:NOPBTD} implies that the expectation values of the output number operators are $\langle\hat{n}_{b_{\_k}}\rangle=n\tau^2+\Gamma^2$ and $\langle\hat{n}_{b_{-\_k}}\rangle=n\Gamma^2+\Gamma^2$. In addition, since the respective photon numbers of the input modes are $\langle\hat{n}_{a_{\_k}}\rangle=n$ and $\langle\hat{n}_{a_{-\_k}}\rangle=0$, the difference in the average number of photons after and before the temporal discontinuity equals to 
\begin{equation}
\langle\hat{n}_{b_{\_k}}+\hat{n}_{b_{-\_k}}\rangle-\langle\hat{n}_{a_{\_k}}+\hat{n}_{a_{-\_k}}\rangle=2\Gamma^2(1+n).
\label{eq:DEXPV}
\end{equation} 
This equation shows that the initial photon number $n$ is not conserved at the time interface, but instead the temporal jump creates on average $2\Gamma^2(1+n)$ additional photons. The nonconservation of the photon number is a manifestation of the Bogoliubov transformation in Eq.~(\ref{eq:bamodes1}) and a consequence of the external work when changing the refractive index. For example, Eq.~(\ref{eq:DEXPV}) states that under the condition of $\Gamma^2=1/[2(n+1)]$, which in terms of the refractive indexes translates into $n_2=[(2+n\pm\sqrt{3+2n})/(1+n)]n_1$, we generate on average a single photon. In this case, if the initial number of photons is zero ($n=0$), the refractive index of the medium after $t=0$ should be $2\pm\sqrt{3}$ times the refractive index of the initial medium. However, although the average photon number is not preserved under the time jump, we note that the average photon number difference between the modes remains invariant:
\begin{equation}
{\langle\hat{n}_{b_{\_k}}-\hat{n}_{b_{-\_k}}\rangle=\langle\hat{n}_{a_{\_k}}-\hat{n}_{a_{-\_k}}\rangle}=n.
\end{equation}

%This property can be understood in terms of Eq.~(\ref{eq:NOPBTD}), according to which photons are created (or annihilated) in pairs.

Having derived the number operators and their expectation values of the output modes, we calculate the corresponding photon number fluctuations in terms of the variances $\Delta n^2=\langle\hat{n}^2\rangle-\langle\hat{n}\rangle^2$. In accordance, after some algebraic steps, we deduce that 
\begin{equation}
\begin{split}
\Delta n_{b_{\_k}}^2&=\Gamma^2\tau^2(n+1)=\Gamma^2(\langle\hat{n}_{b_{\_k}}\rangle+1),\cr
\Delta n_{b_{-\_k}}^2&=\Gamma^2\tau^2(n+1)=\tau^2\langle\hat{n}_{b_{-\_k}}\rangle.
\end{split}
\label{eq:varianceFL}
\end{equation}
The first equalities on the two lines in Eq.~\eqref{eq:varianceFL} reveal that the time interface engenders the same amount of nonzero photon number fluctuations in both output modes, even though the fluctations are strictly zero for the input modes due to the Fock state $\ket{n,0}$. However, and more importantly, the second equalities on the two lines in Eq.~\eqref{eq:varianceFL} underline that the quantum light characters of the forward and backward output modes are drastically different. First, since $\tau>1$, we observe that for the backward mode the variance is always larger than the expectation value. Hence, this mode corresponds to super-Poissonian light. Second, because of the two possibilities $\Gamma<1$ or $\Gamma>1$, we discover that for the forward mode the variance can be smaller or larger than the expectation value. This means that the forward mode after the temporal jump can have the character of either sub-Poissonian or super-Poissonian light. By using the ratio parameter $\beta=\sqrt{n_2/n_1}$ and requiring $(\Delta n_{b_{\_k}}^2/\langle\hat{n}_{b_{\_k}}\rangle)>1$ in Eq.~\eqref{eq:varianceFL}, the condition for having super-Poissonian statistics is
\begin{equation}
\beta>\Big({n\over n+1}\Big)^{1\over4}+\bigg[\Big({n\over n+1}\Big)^{1\over2}+1\bigg]^{1\over2}\quad\mathrm{or}\quad\beta<-\Big({n\over n+1}\Big)^{1\over4}+\bigg[\Big({n\over n+1}\Big)^{1\over2}+1\bigg]^{1\over2}.
\label{eq:CSPSPWAVE}
\end{equation} 
Based on the above expressions, we see that if the initial number of photons is considerable ($n\gg1$), the aforementioned conditions are simplified to $\beta>1+\sqrt{2}$ and $\beta<-1+\sqrt{2}$, in other words $n_2>(3+2\sqrt{2})n_1$ or $n_2<(3-2\sqrt{2})n_1$. In the opposite case, if the initial photon number is zero ($n=0$), Eq.~\eqref{eq:CSPSPWAVE} states that the forward mode displays always super-Poissonian statistics regardless of the ratio parameter $\beta$.

The fact that both of the output modes exhibit super-Poissonian photon statistics when $n=0$ can be understood by considering the two-mode squeezed vacuum state in Eq.~(\ref{eq:vacuum}). By tracing over either the forward or backward mode results exactly in the same mixed state:
\begin{equation}
\hat{\rho}_{b_{\_k}}=\hat{\rho}_{b_{-\_k}}=\frac{1}{\tau^2}\sum_{m=0}^{\infty}\Big(\frac{\Gamma}{\tau}\Big)^{2m}\ket{m}\!\bra{m}.
\label{eq:thermal1}
\end{equation}
On then making the association $(\Gamma/\tau)^2=e^{-\hbar\omega_2/k_\mathrm{B}T}$, where $k_\mathrm{B}$ is the Boltzmann constant and $T$ is the temperature, one observes that Eq.~(\ref{eq:thermal1}) translates into
\begin{equation}
\hat{\rho}_{b_{\_k}}=\hat{\rho}_{b_{-\_k}}=\big(1-e^{-\hbar\omega_2/k_\mathrm{B}T}\big)\sum_{m=0}^{\infty}e^{-m\hbar\omega_2/k_\mathrm{B}T}\ket{m}\!\bra{m}.
\label{eq:thermal2}
\end{equation}
This is the standard expression for a thermal state, with the photon fluctuations obeying super-Poissonian statistics in terms of the Bose--Einstein probability distribution~\cite{agarwal2012quantum,gerry2023introductory}.

%%%%%%%%%%%%%%%%%%%%%%%%%%%%%%%%%%%%%%%%%%%%%%%%%%%%%%%%%%%%%%%%%%%%%%%%%%% 

\begin{table}[!t]
\begin{center}
\begin{tabular}{|p{2.2cm}||p{2.5cm}|p{4.5cm}|p{5cm}|}
 \hline
 \multicolumn{4}{|c|}{Photon Statistics} \\
 \hline
 Input State & Output Mode & Photon Number Fluctuations & Degree of Second-order Coherence \\
 \hline
 Number State & Backward Wave & Super-Poissonian & Bunched \\ 
 \hline
 Number State & Forward Wave  & Super-Poissonian ($\rm{C}_1$) & Bunched ($\rm{C}_1$) \\
 \hline
 Number State & Forward Wave  & Sub-Poissonian ($\rm{C}_2$) & Anti-Bunched ($\rm{C}_2$) \\ 
 \hline 
 Coherent State & Backward Wave & Super-Poissonian & Bunched \\ 
 \hline 
 Coherent State & Forward Wave & Super-Poissonian & Bunched \\ 
 \hline
\end{tabular}
\end{center}
\caption{Photon statistics regarding the forward and backward output modes as a result of the temporal discontinuity at $t=0$. Here, $\rm{C}_1$ refers to the condition that $\beta>\gamma_1$ or $\beta<\gamma_2$ in which $\gamma_1$ and $\gamma_2$ are given by the right-hand sides of the inequalities in Eq.~\eqref{eq:CSPSPWAVE}. On the other hand, $\rm{C}_2$ represents the condition that $\beta<\gamma_1$ and $\beta>\gamma_2$.}
\label{TAB:PSSBA}
\end{table}

\subsection{Degree of second-order coherence}

After determining the super-Poissonian and sub-Poissonian photon statistics of the output modes, it is sensible to examine also the bunched and antibunched photon character of the modes. To this end, we consider the degree of second-order coherence at a single space--time point, which for a generic number operator $\hat{n}$ reads~\cite{fox2006quantum}
\begin{equation}
g^{(2)}(0)=\frac{\langle\hat{n}(\hat{n}-1)\rangle}{\langle\hat{n}\rangle^2}=1+\frac{\Delta n^2-\langle\hat{n}\rangle}{\langle\hat{n}\rangle^2}.
\label{eq:g2}
\end{equation} 
On using the previously calculated expectation values and variances of the number operators in Eq.~(\ref{eq:NOPBTD}), we can determine from Eq.~(\ref{eq:g2}) the conditions under which the photons of the forward and backward modes after the temporal jump exhibit bunching [$g^{(2)}(0)>1$] or antibunching [$g^{(2)}(0)<1$].

After performing the calculations, we end up with
\begin{equation} 
g^{(2)}_{b_{\_k}}(0)=1+{1+(n+1)(\Gamma^4-1)\over (n+\Gamma^2+n\Gamma^2)^2},
\quad g^{(2)}_{b_{-\_k}}(0)=1+{1\over n+1}.
\label{eq:DSOCNS}
\end{equation}
First, we explicitly see that the $g^{(2)}_{b_{-\_k}}(0)$ function for the backward mode does not depend on the variable $\Gamma$. It is a function of only $n$. Second, Eq.~\eqref{eq:DSOCNS} indicates that $1<g^{(2)}_{b_{-\_k}}(0)\leq2$, with the lower and upper limits met when $n\rightarrow\infty$ and $n=0$, respectively. Hence, regardless of the initial photon number $n$, the backward output mode corresponds to bunched light. Note that the condition $g^{(2)}_{b_{-\_k}}(0)=2$ when $n=0$ is a consequence of the fact that in this case the backward mode is in the thermal state given by Eqs.~(\ref{eq:thermal1}) and (\ref{eq:thermal2}). However, concerning the $g^{(2)}_{b_{\_k}}(0)$ function of the forward mode, we see from Eq.~(\ref{eq:DSOCNS}) that it is strongly dependent on both $\Gamma$ and $n$. If $\Gamma=0$, we arrive at $g^{(2)}_{b_{\_k}}(0)=1-1/n<1$. This is expected, because for $\Gamma=0$ there is no temporal discontinuity ($n_2=n_1$), whereupon the forward output mode merges with the forward input mode whose specific antibunched light character arises from the Fock state $\ket{n}$ [see Eq.~(\ref{eq:g2})]. On the other hand, if $\Gamma$ tends to infinity, $g^{(2)}_{b_{\_k}}(0)=1+1/(n+1)>1$. Thus, there is a particular value of $\Gamma$ for which the $g^{(2)}_{b_{\_k}}(0)$ function becomes unity. So depending on the value range of $\Gamma$, the photons in the forward mode can display either antibunching [$g^{(2)}_{b_{\_k}}(0)<1$] or bunching [$g^{(2)}_{b_{\_k}}(0)>1$)]. We have thereby deduced exactly the same conditions for antibunching and bunching that were derived respectively for sub-Poissonian and super-Poissonian statistics in the previous section. This is consistent with the conventional understanding that antibunched light is sub-Poissonian and bunched light is super-Poissonian. We have summarized all these key conclusions explained above about photon statistics in Table~\ref{TAB:PSSBA} (see the first three rows).

%%%%%%%%%%%%%%%%%%%%%%%%%%%%%%%%%%%%%%%%%%%%%%%%%%%%%%%%%%%%%%%%%%%%%%%%%%%
%%%%%%%%%%%%%%%%%%%%%%%%%%%%%%%%%%%%%%%%%%%%%%%%%%%%%%%%%%%%%%%%%%%%%%%%%%%

\subsection{Coherent states} 

In the previous subsections, we studied the situation where the forward input mode is in the number state $\ket{n}$. Here, we concisely look into the scenario in which the forward mode before the temporal discontinuity is in the coherent state
\begin{equation}
\vert\alpha\rangle=e^{-\frac{|\alpha|^2}{2}}\sum_{n=0}^\infty{\alpha^n\over\sqrt{n!}}\vert n\rangle,
\label{eq:coherent}
\end{equation} 
where $\alpha$ is the eigenvalue of the annihilation operator $\hat{a}_{\_k}$. In this case, by using Eqs.~(\ref{eq:NOPBTD}) and (\ref{eq:coherent}), we find that the expectation values of the number operators for the forward and backward output modes are $\langle\hat{n}_{b_{\_k}}\rangle=\tau^2\vert\alpha\vert^2+\Gamma^2$ and $\langle\hat{n}_{b_{-\_k}}\rangle=\Gamma^2(\vert\alpha\vert^2+1)$, respectively. The difference between the average photon numbers before and after the temporal jump is therefore $(\langle\hat{n}_{b_{\_k}}\rangle+\langle\hat{n}_{b_{-\_k}}\rangle)-(\langle\hat{n}_{a_{\_k}}\rangle+\langle\hat{n}_{a_{-\_k}}\rangle)=2\Gamma^2(1+\vert\alpha\vert^2)$, which is similar to Eq.~\eqref{eq:DEXPV} in the context of the number state. 

We continue by deducing the variances of the number operators. This allows us to determine the photon number fluctuations and the photon statistics class (sub-Poissonian or super-Poissonian) of the created modes. Recall that the coherent state is associated with a Poissonian distribution because the variance of the number operator is equal to the mean value. However, as we observed for the number state, the temporal discontinuity may essentially change the situation. By again employing Eqs.~(\ref{eq:NOPBTD}) and (\ref{eq:coherent}), we infer that in the case of an input coherent state
\begin{equation}
\begin{split}
&\Delta n_{b_{\_k}}^2=\Gamma^2\tau^2\Big(\vert\alpha\vert^2+1+{\tau^2\over\Gamma^2}\vert\alpha\vert^2\Big)=\Gamma^2\Big(\langle\hat{n}_{b_{\_k}}\rangle+1+{\tau^4\over\Gamma^2}\vert\alpha\vert^2\Big),\cr
&\Delta n_{b_{-\_k}}^2=\Gamma^2\tau^2\Big(\vert\alpha\vert^2+1+{\Gamma^2\over\tau^2}\vert\alpha\vert^2\Big)=\tau^2\Big(\langle\hat{n}_{b_{-\_k}}\rangle+{\Gamma^4\over\tau^2}\vert\alpha\vert^2\Big).
\end{split}
\label{eq:varVARCS}
\end{equation}
Comparing these results with those in Eq.~\eqref{eq:varianceFL} that we achieved for the number state, we see from the first expressions in Eq.~(\ref{eq:varVARCS}) that this time the photon number variances of the forward and backward modes are not the same. There is instead a difference which equals $\Delta n_{b_{\_k}}^2-\Delta n_{b_{-\_k}}^2=(\tau^2+\Gamma^2)\vert\alpha\vert^2$. From the second expressions in Eq.~(\ref{eq:varVARCS}) we notice that for the backward mode the variance is larger than the expectation value because $\tau>1$. Consequently, it corresponds again to super-Poissonian light. Concerning the forward mode, one might initially think that similarly to the number state it can have sub- or super-Poissonian statistics, because the expression inside the parentheses in Eq.~\eqref{eq:varVARCS} is multiplied by $\Gamma^2$ which can be zero. However, a quick calculation reveals that the variance can be cast into the form $\Delta n_{b_{\_k}}^2=\tau^2(\langle\hat{n}_{b_{\_k}}\rangle+\Gamma^2|\alpha|^2)>1$, which shows that the forward mode is also following super-Poissonian photon statistics.

As a final point, similarly to what we did for the input number state, we deduce the degree of second-order coherence [Eq.~(\ref{eq:g2})] of the output modes when the input state is a coherent state. Having the associated expectation values and the variances, we derive that 
\begin{equation}
g^{(2)}_{b_{{\_k}}}(0)=2-{\tau^4\vert\alpha\vert^4\over(\tau^2\vert\alpha\vert^2+\Gamma^2)^2},\quad 
g^{(2)}_{b_{-{\_k}}}(0)=2-{\vert\alpha\vert^4\over(\vert\alpha\vert^2+1)^2}.
\label{eq:eq:g20eq}
\end{equation} 
Akin to the input number state [Eq.~(\ref{eq:DSOCNS})], we observe that $g^{(2)}_{b_{-{\_k}}}(0)$ of the backward mode in Eq.~(\ref{eq:eq:g20eq}) is completely independent of the material parameters, while for the forward mode $g^{(2)}_{b_{{\_k}}}(0)$ depends on $\tau$ and $\Gamma$. In addition, due to the two latter terms in Eq.~(\ref{eq:eq:g20eq}) we conclude that $1<g^{(2)}_{b_{{\_k}}}(0)\leq2$ and $1<g^{(2)}_{b_{{-\_k}}}(0)\leq2$, which means that both modes correspond to bunched light. Remember that, in contrast, for the input number state the forward output mode can exhibit photon antibunching under some conditions. The main conclusions about the output photon statistics for the input coherent state are summarized in Table~\ref{TAB:PSSBA}.

%%%%%%%%%%%%%%%%%%%%%%%%%%%%%%%%%%%%%%%%%%%%%%%%%%%%%%%%%%%%%%%%%%%%%%%%%%%%
%%%%%%%%%%%%%%%%%%%%%%%%%%%%%%%%%%%%%%%%%%%%%%%%%%%%%%%%%%%%%%%%%%%%%%%%%%%%
%%%%%%%%%%%%%%%%%%%%%%%%%%%%%%%%%%%%%%%%%%%%%%%%%%%%%%%%%%%%%%%%%%%%%%%%%%%%
%%%%%%%%%%%%%%%%%%%%%%%%%%%%%%%%%%%%%%%%%%%%%%%%%%%%%%%%%%%%%%%%%%%%%%%%%%%%

\subsection{Vacuum generation and state discrimination} 

In our previous discussion, we explored the output state when the initial state was set to $\vert n,0\rangle$, demonstrating the remarkable phenomenon of photon-pair generation. This occurrence is primarily attributed to the first exponential term of the two-mode squeeze operator in Eq.~\eqref{eq:TERMs}, which symmetrically incorporates creation operators. Conversely, we see that the third exponential term of the squeeze operator features annihilation operators in its argument, but even so, we did not observe the destruction of photons. The explanation for this lies in our selection of the initial state; we assumed that the incident backward mode is in the vacuum state $\vert0\rangle$. Consequently, this particular term in Eq.~\eqref{eq:TERMs} exerts no influence on the state $\vert n,0\rangle$. However, if the incident backward mode contains photons, the last term in Eq.~\eqref{eq:TERMs} may destroy photons, and thus, the outcome can be drastically different. Indeed, if we apply it on the state $\vert n,\Pi\rangle$ where $0<\Pi\leq n$ and utilize the Taylor expansion, we obtain that
\begin{equation}
e^{-\tanh(r)\hat{K}_{-}}\vert n,\Pi\rangle=\sum_{m=0}^\Pi(-1)^m{\sqrt{n!\Pi!}\over m!\sqrt{(n-m)!(\Pi-m)!}}\tanh^m(r)\vert n-m,\Pi-m\rangle,
\label{eq:S33nP}
\end{equation} 
which explicitly illustrates the phenomenon of photon-pair destruction.

To derive the output state, we must also consider the other terms of the squeeze operator. The action of the second term in Eq.~\eqref{eq:TERMs} on the state $\vert n-m,\Pi-m\rangle$ in Eq.~\eqref{eq:S33nP} results in 
\begin{equation}
e^{\ln[1-\tanh^2(r)]\hat{K}_0}\vert n-m,\Pi-m\rangle=[1-\tanh^2(r)]^{{n+\Pi-2m+1\over2}}\vert n-m,\Pi-m\rangle,
\label{eq:S22nP}
\end{equation} 
indicating that $\vert n-m,\Pi-m\rangle$ is an eigenstate. Next, we must consider the first term of the squeeze operator, whose operation on the state $\vert n-m,\Pi-m\rangle$ yields 
\begin{equation}
e^{\tanh(r)\hat{K}_{+}}\vert n-m,\Pi-m\rangle=\sum_{l=0}^\infty{\sqrt{(n-m+l)!(\Pi-m+l)!}\over l!\sqrt{(n-m)!(\Pi-m)!}}\tanh^l(r)\vert n-m+l,\Pi-m+l\rangle.
\label{eq:S11nP}
\end{equation} 
By eventually combining Eqs.~\eqref{eq:S33nP}, \eqref{eq:S22nP}, and \eqref{eq:S11nP}, we deduce a closed-form solution for the output state which is given by
\begin{equation}
\begin{split}
&\vert\Psi\rangle=\Lambda\sum_{l=0}^\infty\sum_{m=0}^\Pi{(-1)^m\over l!m!}{\sqrt{(n-m+l)!(\Pi-m+l)!}\over(n-m)!(\Pi-m)!}{\tanh^{(m+l)}(r)\over[1-\tanh^2(r)]^{m}}\vert n-m+l,\Pi-m+l\rangle,
\end{split}
\label{EQ:GENPIVINS}
\end{equation}
where $\Lambda=\sqrt{n!\Pi!}[1-\tanh^2(r)]^{n+\Pi+1\over2}$. As a sanity check, we choose $\Pi=0$ and confirm that Eq.~\eqref{EQ:GENPIVINS} reduces to Eq.~\eqref{eq:eq:OSAJ1}, which we discussed earlier. Equation~\eqref{EQ:GENPIVINS}, which provides a general framework, simultaneously encapsulates a combination of two effects: Photon-pair generation and destruction. In particular, the symmetric scenario $n=\Pi$ where $n=m-l$ offers the possibility to generate the vacuum state $\vert0,0\rangle$ within the superposition of Eq.~\eqref{EQ:GENPIVINS}. Since the integer $m$ is equal to or lower than $n$, the only conditions that need to be met are $m=n$ and $l=0$. Accordingly, the state $\vert0,0\rangle$ is produced with a probability $P=[1-\tanh^2(r)]\tanh^{2n}(r)$. 

We also want to highlight another interesting property concealed within Eq.~\eqref{EQ:GENPIVINS}. For simplicity, we consider a symmetric scenario with single-photon illumination in each mode, specifically setting $n=\Pi=1$. In this case, Eq.~\eqref{EQ:GENPIVINS} takes the form 
\begin{equation}
\vert\Psi\rangle=[1-\tanh^2(r)]^{1\over2}\bigg\{-\tanh(r)\vert0,0\rangle+\sum_{l=1}^\infty\tanh^l(r)\Big[{1-\tanh^2(r)\over\tanh(r)}l-\tanh(r)\Big]\vert l,l\rangle\bigg\}.
\label{eq:11UQSDTI}
\end{equation} 
Firstly, we note that our general expression for the probability of the state $\vert0,0\rangle$ aligns with the above equation when $n=1$. More importantly, we observe that the coefficient inside the square brackets in the series is proportional to the integer $l$. This observation opens up the possibility of eliminating a specific state from the superposition in Eq.~\eqref{eq:11UQSDTI}. To achieve such an elimination, we require the following criterion to hold:
\begin{equation}
\tanh^2(r)=\Big({\Gamma\over\tau}\Big)^2={l\over l+1}, 
\end{equation}
which in terms of the refractive indices translates to two possible relationships: $n_2/n_1=\big[1+\sqrt{l/(l+1)}\big]/\big[1-\sqrt{l/(l+1)}\big]$ or $n_2/n_1=\big[1-\sqrt{l/(l+1)}\big]/\big[1+\sqrt{l/(l+1)}\big]$. For instance, to eliminate the state $\vert1,1\rangle$, the refractive index ratio should be either $n_2/n_1=5.8$ or $n_2/n_1=0.17$. By designing the time interface to match one of these specific values, we ensure that repeated measurements will never yield a scenario where one photon is detected in the forward direction and another single photon in the backward direction. This phenomenon represents a form of unambiguous state discrimination.

%%%%%%%%%%%%%%%%%%%%%%%%%%%%%%%%%%%%%%%%%%%%%%%%%%%%%%%%%%%%
%%%%%%%%%%%%%%%%%%%%%%%%%%%%%%%%%%%%%%%%%%%%%%%%%%%%%%%%%%%%
%%%%%%%%%%%%%%%%%%%%%%%%%%%%%%%%%%%%%%%%%%%%%%%%%%%%%%%%%%%%
%%%%%%%%%%%%%%%%%%%%%%%%%%%%%%%%%%%%%%%%%%%%%%%%%%%%%%%%%%%%

\subsection{Quantum state freezing} 

As indicated by Eqs.~\eqref{eq:Soperator} and \eqref{EQUT:EM12NOVUTC}, when the squeezing parameter $r$ is zero, the two-mode squeeze operator reduces to the identity operator and the output mode operators become identical to the input mode operators. Consequently, the resulting output state also remains unaltered, whereupon we henceforth refer to such a scenario as quantum state freezing. According to Eq.~\eqref{EQEQEQRRR}, the condition $r=0$ implies that $\Gamma=0$, which in turn necessitates that the refractive indices $n_1$ and $n_2$ are equal. This equality, however, negates the very existence of a time interface, which is based on a change in refractive index. The simultaneous existence of a time interface and the freezing of the quantum state thus requires an additional degree of freedom. In the context of isotropic and nondispersive materials, an additional degree of freedom is provided by the material's relative permeability, a parameter hitherto unaccounted for in our analysis.

Assume that the relative permeability is not unity. This assumption does not affect the electric field operators in Eq.~\eqref{EQ:EQBAVFOEVBAV}, and consequently, the flux density operators remain unchanged. However, the implications of the boundary conditions become significantly different. The relationship between frequencies in the two media is now given by $\omega_2=\omega_1\sqrt{\mu_1\epsilon_1/\mu_2\epsilon_2}$, where $\mu$ represents the permeability. Applying this relationship and imposing the boundary conditions, the two parameters $\tau$ and $\Gamma$ in Eq.~\eqref{eq:taugamma} are modified as 
\begin{equation}
\tau=\frac{1}{2}\Big(\sqrt{\frac{\eta_1}{\eta_2}}+\sqrt{\frac{\eta_2}{\eta_1}}\Big),\quad 
\Gamma=\frac{1}{2}\Big(\sqrt{\frac{\eta_1}{\eta_2}}-\sqrt{\frac{\eta_2}{\eta_1}}\Big).
\label{EQEQ:RVMGT}
\end{equation}
Here, $\eta=\sqrt{\mu/\epsilon}$ is the wave impedance of each medium. Equation~\eqref{EQEQ:RVMGT} reveals that if $\eta_2=\eta_1$, the parameter $\Gamma$ vanishes ($\tau=1$), resulting in a zero squeezing parameter according to Eq.~\eqref{EQEQEQRRR}. Hence, the initial quantum state is preserved. On the other hand, and crucially, the equality of wave impedances does not preclude the existence of a time interface, as the angular frequency can still undergo a shift. If $\mu_2=\kappa\mu_1$ ($\kappa$ is a real number), $\eta_2=\eta_1$ requires that $\epsilon_2=\kappa\epsilon_1$, which leads to the frequency transformation $\omega_2 = \omega_1/\kappa$. Depending on the value of $\kappa$, the frequency may be shifted upwards or downwards. The above analysis unveils a remarkable phenomenon: We can achieve frequency conversion without altering the quantum state of the system.

%%%%%%%%%%%%%%%%%%%%%%%%%%%%%%%%%%%%%%%%%%%%%%%%%%%%%%%%%%%%%%%%%%%%%%%%%%%
%%%%%%%%%%%%%%%%%%%%%%%%%%%%%%%%%%%%%%%%%%%%%%%%%%%%%%%%%%%%%%%%%%%%%%%%%%%
%%%%%%%%%%%%%%%%%%%%%%%%%%%%%%%%%%%%%%%%%%%%%%%%%%%%%%%%%%%%%%%%%%%%%%%%%%%
%%%%%%%%%%%%%%%%%%%%%%%%%%%%%%%%%%%%%%%%%%%%%%%%%%%%%%%%%%%%%%%%%%%%%%%%%%%

\subsection{Possible experimental realization} 

To validate the theoretical results presented in this paper, we propose an experimental approach using a system that precisely fulfills our assumption of nondispersive materials: A transmission line operating in the microwave regime. Transmission lines are well-established nondispersive systems widely employed in radio engineering for energy transfer at a constant velocity of $1/\sqrt{LC}$, where $L$ and $C$ represent the inductance and capacitance per unit length, respectively~\cite[Chapter~2]{Pozar2012Microwave}. To minimize dissipation and effectively eliminate thermal excitations, the experiment should be conducted at ultra-low temperatures, specifically in the millikelvin range. This cryogenic environment enables the use of a superconducting transmission line, typically operating in the 4--8 GHz frequency range at temperatures around 10~mK. For the experimental setup, we suggest employing a planar transmission line with a characteristic impedance of 50~$\Omega$, which is standard in low- and high-temperature microwave applications~\cite{Pozar2012Microwave}. 

\begin{figure}[!t]
\centerline{\includegraphics[width=1\columnwidth]{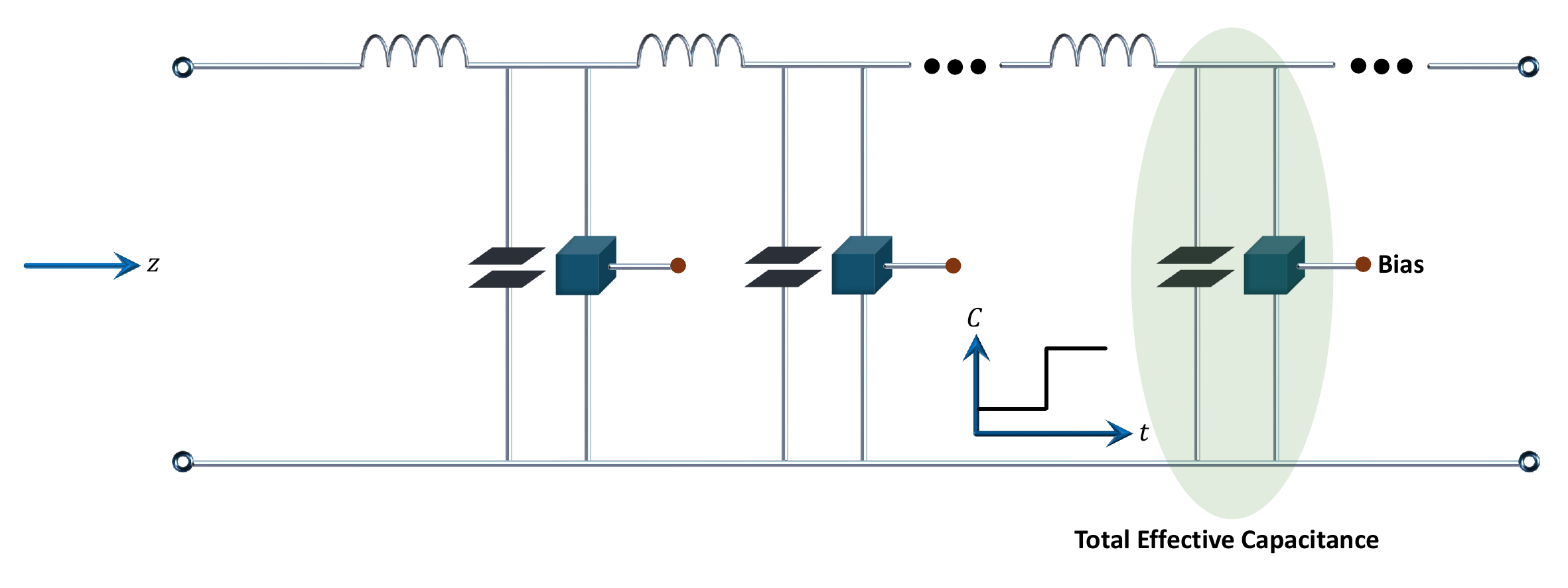}}
\caption{A superconducting transmission line which is designed to work in the microwave regime and at ultra-low temperatures. The line includes electronic elements modelled as boxes and connected periodically along the line. The equivalent capacitance provided by each element is parallel with the intrinsic capacitance of the line. This equivalent capacitance is tuned based on an external bias.}
\label{fig:SCONTLEVERT}
\end{figure}

A key requirement of the experiment is that the capacitance (per unit length) of the line must change instantaneously at a specific moment. To achieve this, we propose using lumped electronic components periodically distributed along the line, providing an equivalent capacitance. These components should be positioned such that their equivalent capacitance is in parallel with the line's intrinsic capacitance (see Fig.~\ref{fig:SCONTLEVERT}). Consequently, the total effective capacitance per unit length is the sum of the line's capacitance and the capacitance provided by the electronic elements. The capacitance of these electronic elements can be tuned, for example, by an external voltage bias as shown in Fig.~\ref{fig:SCONTLEVERT}. Such tuning capability allows the total effective capacitance of the line to change value at a specific moment. It is crucial to emphasize that the charge associated with the capacitance at the moment of the jump must remain continuous. In other words, we must ensure that no additional charges are introduced to the system during this transition.

Regarding such a transmission line, in the context of circuit quantum electrodynamics, the flux and charge operators for the forward and backward modes can be expressed as~\cite{blais2021circuit}
\begin{equation}
\begin{split}
&\hat{\Phi}(z,t)=\Big(\frac{\hbar}{4\pi C}\Big)^{1/2}\sum_{k^\prime}\frac{1}{\sqrt{\omega}}[\hat{a}_{k^\prime}e^{i(k^\prime z-\omega t)}+\hat{a}^\dagger_{k^\prime}e^{-i(k^\prime z-\omega t)}],\cr
&\hat{Q}(z,t)=-i\Big(\frac{\hbar C}{4\pi}\Big)^{1/2}\sum_{k^\prime}\sqrt{\omega}[\hat{a}_{k^\prime}e^{i(k^\prime z-\omega t)}-\hat{a}^\dagger_{k^\prime}e^{-i(k^\prime z-\omega t)}]. 
\end{split}
\label{EQEQFLUXCHAM}
\end{equation} 
It is important to note that transmission lines lack polarization and are one-dimensional systems. Consequently, the phase constant $k^\prime\in\{k,-k\}$ is a scalar quantity, where $k=\omega\sqrt{LC}$. We suppose that the energy is transferred along the $z$ direction. Equation~\eqref{EQEQFLUXCHAM} is applicable both before and after the temporal jump using the proper values for the capacitance and angular frequency and also using the proper mode operators (i.e., $\hat{a}_{k^\prime}$ before the jump and $\hat{b}_{k^\prime}$ after the jump). To maintain finite voltage and electric current, the flux and charge operators must be continuous across the transition. This continuity requirement is analogous to the continuity of magnetic and electric flux densities in bulk media and serves as our boundary conditions. By imposing these conditions, we derive the analogous transformation as described by Eq.~\eqref{eq:bamodes1} for bulk media. The key distinction lies in replacing the refractive index with the square root of the capacitance. In other words, now, we obtain
\begin{equation}
\tau=\frac{1}{2}\Big(\sqrt[4]{\frac{C_2}{C_1}}+\sqrt[4]{\frac{C_1}{C_2}}\Big),\quad 
\Gamma=\frac{1}{2}\Big(\sqrt[4]{\frac{C_2}{C_1}}-\sqrt[4]{\frac{C_1}{C_2}}\Big).
\end{equation} 
The parameters $C_1$ and $C_2$ correspond to the total effective capacitance (per unit length)  before and after the transition, respectively. This formulation suggests that transmission lines can effectively substitute bulk media to experimentally verify our theoretical predictions. However, as mentioned, such a setup necessitates the use of superconducting transmission lines at low temperatures.

%%%%%%%%%%%n the context of%%%%%%%%%%%%%%%%%%%%%%%%%%%%%%%%%%%%%%%%%%%%% Section 3
%%%%%%%%%%%%%%%%%%%%%%%%%%%%%%%%%%%%%%%%%%%%%%%%%%%%%%%% 

\section{Conclusions} 
\label{sec:conclusion}

In summary, we have conducted a detailed investigation of the scattering of quantized electromagnetic radiation from a time interface between two isotropic and nondispersive media. First, from a classical electrodynamics point of view, we briefly reviewed the main features of wave scattering from such an interface. Subsequently, as the principal aim, we studied the scattering properties from a quantum optics perspective. By considering the situation where a forward and a backward propagating mode are present in the initial medium, we inspected the transformation of the bosonic mode operators due to the existence of a temporal discontinuity in the refractive index. It was specifically shown that the temporal boundary results in a unitary evolution of the mode operators and the related quantum states that is governed by the two-mode squeeze operator. Taking the initial forward and backward modes to be in number and vacuum states, respectively, we then observed that photon-pair generation is an intrinsic characteristic of the time interface. We further contemplated the associated photon probability distribution and analyzed the conditions of creating a one-photon pair, with the maximum probability being in the range 25--36$\%$ depending on the incident photon number and the refractive-index ratio. In addition, for two different input states of the forward mode, a number state and a coherent state, we scrutinized the photon number fluctuations and the degree of second-order coherence of the individual output modes to assess their photon statistics. We saw that the backward mode corresponds always to super-Poissonian (bunched) light after the temporal discontinuity, while for the number state, unlike with the coherent state, the forward mode can represent 
sub-Poissonian (antibunched) light. 

Moreover, our investigation indicated that interesting phenomena emerge when challenging preliminary assumptions. For instance, we demonstrated that photon-pair destruction occurs when the initial backward propagating mode is not in a vacuum state but in a number state. In a symmetric scenario where equal numbers of photons are present in the initial modes, photon-pair destruction may result in the generation of a vacuum state $\vert0,0\rangle$. The inclusion of photons in the initial backward propagating mode also gives rise to an unambiguous discrimination effect, effectively negating one state in the output superposition. Another significant example is that when the material under study possesses both magnetic and electric responses (i.e., relative permeability is not unity anymore), it becomes possible to match the impedances of the two media on either side of the interface. This impedance matching, characterized by a zero squeezing parameter, leads to the preservation of the initial quantum state across the interface while the angular frequency is shifted. Finally, within the framework of circuit quantum electrodynamics, we proposed a method for experimentally verifying our theoretical predictions. This method uses a superconducting transmission line operating at extremely low temperatures, a setup that is feasible and already employed in quantum computing applications.

The analysis presented in this paper supports further theoretical investigation into more sophisticated time-interface mechanisms. Through the lens of classical electrodynamics, we know that if the dielectric material under study becomes dispersive, and temporal nonlocality plays a role, the time interface results in the generation of at least two angular frequencies, in contrast to the nondispersive scenario. This feature is anticipated to bring uncharted quantum optical effects and possibilities for further careful examination. Another enticing future problem is to expand our current study to temporal slabs, which in turn paves the way towards advancing a rigorous quantum theory of field interaction with materials whose corresponding band structure shows wavevector gaps due to the periodic modulation of the refractive index.

%%%%%%%%%%%%%%%%%%%%%%%%%%%%%%%%%%%%%%%%%%%%%%%%%%%%%%%% 

\section*{Acknowledgment}
This work was supported by the Research Council of Finland (Grants Nos.~354918, 346518, 349396, and 336119). M.S.M.~would like to express his gratitude to Martti Hanhisalo for useful general discussions. Additionally, M.S.M.~is thankful to Sorin Paraoanu for his valuable insights regarding low-temperature superconducting transmission lines.

%%%%%%%%%%%%%%%%%%%%%%%%%%%%%%%%%%%%%%%%%%%%%%%%%%%%%%%%  

\bibliographystyle{IEEEtran}
\bibliography{IEEEabrv,references}

\begin{thebibliography}{10}
\providecommand{\url}[1]{#1}
\csname url@samestyle\endcsname
\providecommand{\newblock}{\relax}
\providecommand{\bibinfo}[2]{#2}
\providecommand{\BIBentrySTDinterwordspacing}{\spaceskip=0pt\relax}
\providecommand{\BIBentryALTinterwordstretchfactor}{4}
\providecommand{\BIBentryALTinterwordspacing}{\spaceskip=\fontdimen2\font plus
\BIBentryALTinterwordstretchfactor\fontdimen3\font minus \fontdimen4\font\relax}
\providecommand{\BIBforeignlanguage}[2]{{%
\expandafter\ifx\csname l@#1\endcsname\relax
\typeout{** WARNING: IEEEtran.bst: No hyphenation pattern has been}%
\typeout{** loaded for the language `#1'. Using the pattern for}%
\typeout{** the default language instead.}%
\else
\language=\csname l@#1\endcsname
\fi
#2}}
\providecommand{\BIBdecl}{\relax}
\BIBdecl

\bibitem{galiffi2022photonics}
E.~Galiffi, R.~Tirole, S.~Yin, H.~Li, S.~Vezzoli, P.~A. Huidobro, M.~G. Silveirinha, R.~Sapienza, A.~Al{\`u}, and J.~Pendry, ``Photonics of time-varying media,'' \emph{Advanced Photonics}, vol.~4, no.~1, pp. 014\,002--014\,002, 2022.

\bibitem{Ptitcyn2023Tutorial}
G.~Ptitcyn, M.~S. Mirmoosa, A.~Sotoodehfar, and S.~A. Tretyakov, ``A tutorial on the basics of time-varying electromagnetic systems and circuits: Historic overview and basic concepts of time-modulation.'' \emph{IEEE Antennas and Propagation Magazine}, 2023.

\bibitem{engheta2023four}
N.~Engheta, ``Four-dimensional optics using time-varying metamaterials,'' \emph{Science}, vol. 379, no. 6638, pp. 1190--1191, 2023.

\bibitem{Engheta20NPH}
------, ``Metamaterials with high degrees of freedom: space, time, and more,'' \emph{Nanophotonics}, vol.~10, no.~1, pp. 639--642, 2021.

\bibitem{morgenthaler1958velocity}
F.~R. Morgenthaler, ``Velocity modulation of electromagnetic waves,'' \emph{IRE Transactions on Microwave Theory and Techniques}, vol.~6, no.~2, pp. 167--172, 1958.

\bibitem{mendoncca2002time}
J.~Mendon{\c{c}}a and P.~Shukla, ``Time refraction and time reflection: two basic concepts,'' \emph{Physica Scripta}, vol.~65, no.~2, p. 160, 2002.

\bibitem{Agrawal2014RTC}
Y.~Xiao, D.~N. Maywar, and G.~P. Agrawal, ``Reflection and transmission of electromagnetic waves at a temporal boundary,'' \emph{Optics letters}, vol.~39, no.~3, pp. 574--577, 2014.

\bibitem{Wilks1988TH}
S.~C. Wilks, J.~M. Dawson, and W.~B. Mori, ``Frequency up-conversion of electromagnetic radiation with use of an overdense plasma,'' \emph{Phys. Rev. Lett.}, vol.~61, pp. 337--340, Jul 1988.

\bibitem{Yugami2002EXPER}
N.~Yugami, T.~Niiyama, T.~Higashiguchi, H.~Gao, S.~Sasaki, H.~Ito, and Y.~Nishida, ``Experimental observation of short-pulse upshifted frequency microwaves from a laser-created overdense plasma,'' \emph{Phys. Rev. E}, vol.~65, p. 036505, Mar 2002.

\bibitem{nishida2012EXP}
A.~Nishida, N.~Yugami, T.~Higashiguchi, T.~Otsuka, F.~Suzuki, M.~Nakata, Y.~Sentoku, and R.~Kodama, ``Experimental observation of frequency up-conversion by flash ionization,'' \emph{Applied Physics Letters}, vol. 101, no.~16, 2012.

\bibitem{water}
V.~Bacot, M.~Labousse, A.~Eddi, M.~Fink, and E.~Fort, ``Time reversal and holography with spacetime transformations,'' \emph{Nature Physics}, vol.~12, pp. 972--977, 2016.

\bibitem{Alu_TL}
H.~Moussa, G.~Xu, S.~Yin, E.~Galiffi, Y.~Ra’di, and A.~Alù, ``Observation of temporal reflection and broadband frequency translation at photonic time interfaces,'' \emph{Nature Physics}, vol.~19, pp. 863--868, 2023.

\bibitem{Segev8PTC}
E.~Lustig, Y.~Sharabi, and M.~Segev, ``Topological aspects of photonic time crystals,'' \emph{Optica}, vol.~5, no.~11, pp. 1390--1395, 2018.

\bibitem{ramaccia2020light}
D.~Ramaccia, A.~Toscano, and F.~Bilotti, ``Light propagation through metamaterial temporal slabs: reflection, refraction, and special cases,'' \emph{Optics Letters}, vol.~45, no.~20, pp. 5836--5839, 2020.

\bibitem{pacheco2020anti}
V.~Pacheco-Pe{\~n}a and N.~Engheta, ``Antireflection temporal coatings,'' \emph{Optica}, vol.~7, no.~4, pp. 323--331, 2020.

\bibitem{akbarzadeh2018inverse}
A.~Akbarzadeh, N.~Chamanara, and C.~Caloz, ``Inverse prism based on temporal discontinuity and spatial dispersion,'' \emph{Optics Letters}, vol.~43, no.~14, pp. 3297--3300, 2018.

\bibitem{pacheco2020aiming}
V.~Pacheco-Pe{\~n}a and N.~Engheta, ``Temporal aiming,'' \emph{Light: Science \& Applications}, vol.~9, no.~1, p. 129, 2020.

\bibitem{xu2021complete}
J.~Xu, W.~Mai, and D.~H. Werner, ``Complete polarization conversion using anisotropic temporal slabs,'' \emph{Optics Letters}, vol.~46, no.~6, pp. 1373--1376, 2021.

\bibitem{mostafa2023spin}
M.~H.~M. Mostafa, M.~S. Mirmoosa, and S.~A. Tretyakov, ``Spin-dependent phenomena at chiral temporal interfaces,'' \emph{Nanophotonics}, vol.~12, no.~14, pp. 2881--2889, 2023.

\bibitem{RizzaCastaldiGaldi23}
C.~Rizza, G.~Castaldi, and V.~Galdi, ``Spin-controlled photonics via temporal anisotropy,'' \emph{Nanophotonics}, vol.~12, no.~14, pp. 2891--2904, 2023.

\bibitem{mirmoosa2023timeIIBM}
M.~S. Mirmoosa, M.~H. Mostafa, A.~Norrman, and S.~A. Tretyakov, ``Time interfaces in bianisotropic media,'' \emph{arXiv:2306.17522}, 2023.

\bibitem{wang2023controlling}
X.~Wang, M.~S. Mirmoosa, and S.~A. Tretyakov, ``Controlling surface waves with temporal discontinuities of metasurfaces,'' \emph{Nanophotonics}, vol.~12, no.~14, pp. 2813--2822, 2023.

\bibitem{Grap19SPP}
A.~V. Shirokova, A.~V. Maslov, and M.~I. Bakunov, ``Scattering of surface plasmons on graphene by abrupt free-carrier generation,'' \emph{Phys. Rev. B}, vol. 100, p. 045424, Jul 2019.

\bibitem{mendoncca2000quantum}
J.~Mendon{\c{c}}a, A.~Guerreiro, and A.~M. Martins, ``Quantum theory of time refraction,'' \emph{Physical Review A}, vol.~62, no.~3, p. 033805, 2000.

\bibitem{SecondRVAgree1}
J.~T. Mendon{\c{c}}a, \emph{Theory of photon acceleration}.\hskip 1em plus 0.5em minus 0.4em\relax CRC Press, 2000.

\bibitem{SecondRVAgree2}
J.~Mendon{\c{c}}a, A.~Martins, and A.~Guerreiro, ``Temporal beam splitter and temporal interference,'' \emph{Physical Review A}, vol.~68, no.~4, p. 043801, 2003.

\bibitem{vazquez2022shaping}
J.~E. V{\'a}zquez-Lozano and I.~Liberal, ``Shaping the quantum vacuum with anisotropic temporal boundaries,'' \emph{Nanophotonics}, vol.~12, no.~3, pp. 539--548, 2022.

\bibitem{liberal2023quantum}
I.~Liberal, J.~E. V{\'a}zquez-Lozano, and V.~Pacheco-Pe{\~n}a, ``Quantum antireflection temporal coatings: quantum state frequency shifting and inhibited thermal noise amplification,'' \emph{Laser \& Photonics Reviews}, vol.~17, no.~9, p. 2200720, 2023.

\bibitem{RVNotAgree1}
S.~A. Horsley and J.~B. Pendry, ``Quantum electrodynamics of time-varying gratings,'' \emph{Proceedings of the National Academy of Sciences}, vol. 120, no.~36, p. e2302652120, 2023.

\bibitem{RVNotAgree2}
J.~Pendry and S.~A. Horsley, ``Qed in space--time varying materials,'' \emph{APL Quantum}, vol.~1, no.~2, 2024.

\bibitem{RVNotAgree3}
A.~Ganfornina-Andrades, J.~E. V{\'a}zquez-Lozano, and I.~Liberal, ``Quantum vacuum amplification in time-varying media with arbitrary temporal profiles,'' \emph{arXiv preprint arXiv:2312.13315}, 2023.

\bibitem{RVNotAgree4}
J.~Echave-Sustaeta, F.~J. Garc{\'\i}a-Vidal, and P.~Huidobro, ``Photon squeezing in photonic time crystals,'' \emph{arXiv preprint arXiv:2405.05043}, 2024.

\bibitem{cheng1989field}
D.~K. Cheng \emph{et~al.}, \emph{Field and wave electromagnetics}.\hskip 1em plus 0.5em minus 0.4em\relax Pearson Education India, 1989.

\bibitem{loudon2000quantum}
R.~Loudon, \emph{The quantum theory of light}.\hskip 1em plus 0.5em minus 0.4em\relax Oxford University Press, 2000.

\bibitem{mirmoosa2022dipole}
M.~Mirmoosa, T.~Koutserimpas, G.~Ptitcyn, S.~Tretyakov, and R.~Fleury, ``Dipole polarizability of time-varying particles,'' \emph{New Journal of Physics}, vol.~24, no.~6, p. 063004, 2022.

\bibitem{leonhardt2010essential}
U.~Leonhardt, \emph{Essential quantum optics: from quantum measurements to black holes}.\hskip 1em plus 0.5em minus 0.4em\relax Cambridge University Press, 2010.

\bibitem{agarwal2012quantum}
G.~S. Agarwal, \emph{Quantum optics}.\hskip 1em plus 0.5em minus 0.4em\relax Cambridge University Press, 2012.

\bibitem{gerry2023introductory}
C.~C. Gerry and P.~L. Knight, \emph{Introductory quantum optics}.\hskip 1em plus 0.5em minus 0.4em\relax Cambridge university press, 2023.

\bibitem{sakurai2020modern}
J.~J. Sakurai and J.~Napolitano, \emph{Modern quantum mechanics}.\hskip 1em plus 0.5em minus 0.4em\relax Cambridge University Press, 2020.

\bibitem{wodkiewicz1985coherent}
K.~Wodkiewicz and J.~Eberly, ``Coherent states, squeezed fluctuations, and the su(2) and su(1, 1) groups in quantum-optics applications,'' \emph{JOSA B}, vol.~2, no.~3, pp. 458--466, 1985.

\bibitem{gilles1992non}
L.~Gilles and P.~Knight, ``Non-classical properties of two-mode su(1, 1) coherent states,'' \emph{Journal of Modern Optics}, vol.~39, no.~7, pp. 1411--1440, 1992.

\bibitem{GriffithsIQM}
D.~J. Griffiths, \emph{Introduction to Quantum Mechanics}.\hskip 1em plus 0.5em minus 0.4em\relax Cambridge University Press, 2017.

\bibitem{barnett1989entropy}
S.~Barnett and S.~Phoenix, ``Entropy as a measure of quantum optical correlation,'' \emph{Physical Review A}, vol.~40, no.~5, p. 2404, 1989.

\bibitem{barnett1991information}
S.~M. Barnett and S.~J. Phoenix, ``Information theory, squeezing, and quantum correlations,'' \emph{Physical Review A}, vol.~44, no.~1, p. 535, 1991.

\bibitem{fox2006quantum}
A.~M. Fox, \emph{Quantum optics: an introduction}.\hskip 1em plus 0.5em minus 0.4em\relax Oxford University Press, USA, 2006, vol.~15.

\bibitem{Pozar2012Microwave}
D.~M. Pozar, \emph{Microwave engineering}.\hskip 1em plus 0.5em minus 0.4em\relax John Wiley \& Sons, 2012.

\bibitem{blais2021circuit}
A.~Blais, A.~L. Grimsmo, S.~M. Girvin, and A.~Wallraff, ``Circuit quantum electrodynamics,'' \emph{Reviews of Modern Physics}, vol.~93, no.~2, p. 025005, 2021.

\end{thebibliography}

\end{document}